\documentclass[aip,
nofootinbib,
rsi,%
reprint,
twocolumn,
longbibliography,
author-numerical%
]{revtex4-2}

\usepackage{graphicx}
\usepackage{textcomp}
\usepackage{natbib}

\usepackage{amsfonts}
\usepackage{amssymb}
\usepackage{amsmath}
\usepackage[usenames]{color}

\usepackage{setspace}
\usepackage{changes}

\newcommand{\pdr }[2]{\dfrac{\partial {#1}}{\partial {#2}}}

\newcommand{\pddr}[2]{\dfrac{\partial^2{#1}}{\partial{#2}^2}}

\newcommand{\pdra }[2]{{\partial  {#1}}/{\partial {#2}}}

\newcommand{\tx}{\tilde{x}}

\newcommand{\tz}{\tilde{z}}

\newcommand{\tit}{\tilde{t}}

\newcommand{\tj}{\tilde{j}}

\newcommand{\tc}{\tilde{c}}

\newcommand{\teta}{\tilde{\eta}}

\newcommand{\tJ}{\tilde{J}}

\newcommand{\tZ}{\tilde{Z}}
\newcommand{\tD}{\tilde{D}}

\newcommand{\tl}{\tilde{l}}

\newcommand{\expo}{{{\rm e}^{\teta_0}}}
\newcommand{\expot}{{{\rm e}^{2\teta_0}}}

\newcommand{\cref }{c_h^{in}}

\newcommand{\tom  }{\tilde{\omega}}

\newcommand{\jlim}{j_{\lim}}

\newcommand{\Cdl}{C_{dl}}

\newcommand{\tjlim}{\tilde{j}_{\lim}}

\newcommand{\lam }{\lambda}

\newcommand{\lcat}{l_t}

\newcommand{\ri}{{\rm i}}

\newcommand{\sqa}{\ri\tom/\tD_b}

\newcommand{\lexp}[1]{\exp\left(#1\right)}
\newcommand{\lnl }[1]{\ln \left(#1\right)}

\newcommand{\etal}{et al.{ }}

\renewcommand{\Im}[1]{\operatorname{Im}\left(#1\right)}

\begin{document}

\sf

\title{Analytical impedance of oxygen transport
      in the channel and gas diffusion layer of a PEM fuel cell}

\author{Andrei Kulikovsky}
\thanks{ECS Active member}
\email{A.Kulikovsky@fz-juelich.de}

\affiliation{Forschungszentrum J\"ulich GmbH   \\
Theory and Computation of Energy Materials (IEK--13)   \\
Institute of Energy and Climate Research,     \\
D--52425 J\"ulich, Germany
}

\altaffiliation[Also at: ]{Lomonosov Moscow State University,
                Research Computing Center, 119991 Moscow, Russia}

\date{\today}

\begin{abstract}
Analytical model for impedance of oxygen transport
in the gas--diffusion layer (GDL) and cathode channel of a PEM fuel cell is developed.
The model is based on transient oxygen mass conservation equations coupled
to the proton current conservation equation in the catalyst layer.
Analytical formula for the ``GDL+channel'' impedance is derived
assuming that the oxygen and proton transport in the cathode catalyst layer (CCL) are fast.
In the Nyquist plot, the resulting impedance consists of two arcs
describing oxygen transport in the air channel (low--frequency arc) and
in the GDL. The characteristic frequency of GDL arc depends
on the CCL thickness: large CCL thickness strongly lowers this frequency.
At small CCL thickness, the high--frequency feature on the arc shape forms.
This effect is important for identification of peaks in
distribution of relaxation times spectra of low--Pt PEMFCs.
\end{abstract}


\maketitle

\section{Introduction}

Electrochemical impedance spectroscopy (EIS)
provides invaluable information on transport properties of PEM fuel cell
in a current production mode, without interruption of cell functioning\cite{Lasia_book_14}.
Not surprisingly, EIS of PEM fuel cells is a rapidly growing field\cite{Tang_20}.
Understanding impedance spectra requires modeling.
Strong criticism of equivalent circuit approach
has been published by Macdonald in his seminal paper\cite{Macdonald_06} and in recent years,
physics--based models for PEMFC impedance tend to replace equivalent
circuit modeling (see recent reviews of Tang \etal\cite{Tang_20}
and Huang \etal\cite{Huang_20}).

Every transport and kinetic process in a PEM fuel cell has its own resonance
frequency. If these frequencies do not overlap, one could identify them
using the distribution of relaxation times  (DRT) technique\cite{Fuoss_41,Schlichlein_02,Effendy_20}.
In addition, DRT analysis of impedance spectra returns
the contribution of every process into the total differential resistance
of the cell. However, correct identification of DRT peaks
is a non--trivial task requiring modeling and experimental work.
Analytical models predicting characteristic frequencies of oxygen
transport processes in the cell could be very helpful in this respect.

Oxygen reduction reaction (ORR) is usually responsible for
a large part of potential loss in a PEMFC.
Oxygen is transported to the catalyst sites through the cathode channel
and gas--diffusion layer; both the transport processes have their signatures in the EIS
spectra. After pioneering experimental
work of Schneider \etal\cite{Ingo_07a,Ingo_07b},
Kulikovsky and Shamardina\cite{Kulikovsky_12f,Kulikovsky_15g},
Maranzana \etal\cite{Maranzana_12} and Chevalier \etal\cite{Chevalier_16b}
developed numerical and analytical models incorporating channel impedance.
Formulas for pure channel impedance have been obtained\cite{Chevalier_16b,Kulikovsky_19b}
assuming fast  oxygen transport through the GDL and cathode catalyst layer (CCL).
However, the coupling between the channel and GDL impedance remained poorly understood.
Recently, Cruz--Manzo and Greenwood reported analytical model for
the GDL+channel impedance\cite{Cruz_Manzo_21}. However, their result missing important effect
of double layer charging on this impedance, as discussed below.

In this work, we develop analytical model for the GDL+channel impedance $Z_{gdlc}$
in a PEM fuel cell. Assuming fast oxygen and proton transport in the CCL,
analytical expression for $Z_{gdlc}$ is derived.
We show that for typical PEMFC parameters, the Nyquist spectrum of $Z_{gdlc}$
consists of two arcs corresponding to oxygen transport
in the channel and GDL. GDL impedance differs from the Warburg finite--length impedance
due to ``non--Warburg'' factor depending on the superficial double
layer capacitance of the electrode. For typical PEMFC parameters,
the characteristic frequency $f_{gdl}$ of the  GDL impedance $Z_{gdl}$ is close to the
Warburg finite--length frequency; however, for larger CCL thickness typical
for non--Pt cells, the non--Warburg  factor strongly lowers $f_{gdl}$.
In the opposite limit of small catalyst layer thickness,
a high--frequency feature on the shape of imaginary part of $Z_{gdl}$
vs frequency forms. Analysis shows that the channel impedance also depends
on the double layer capacitance meaning that impedance
of all oxygen transport medias in the PEMFC ``feel'' double layer
capacitance of the electrode where the oxygen is transported to.
The goal of this paper is to clarify the situation with oxygen transport impedance
in the GDL and channel after interesting and useful, but incomplete
and rather difficult for understanding work of Cruz-Manzo and Greenwood\cite{Cruz_Manzo_21}.

\section{Model}

\begin{figure}
	\begin{center}
		\includegraphics[scale=0.7]{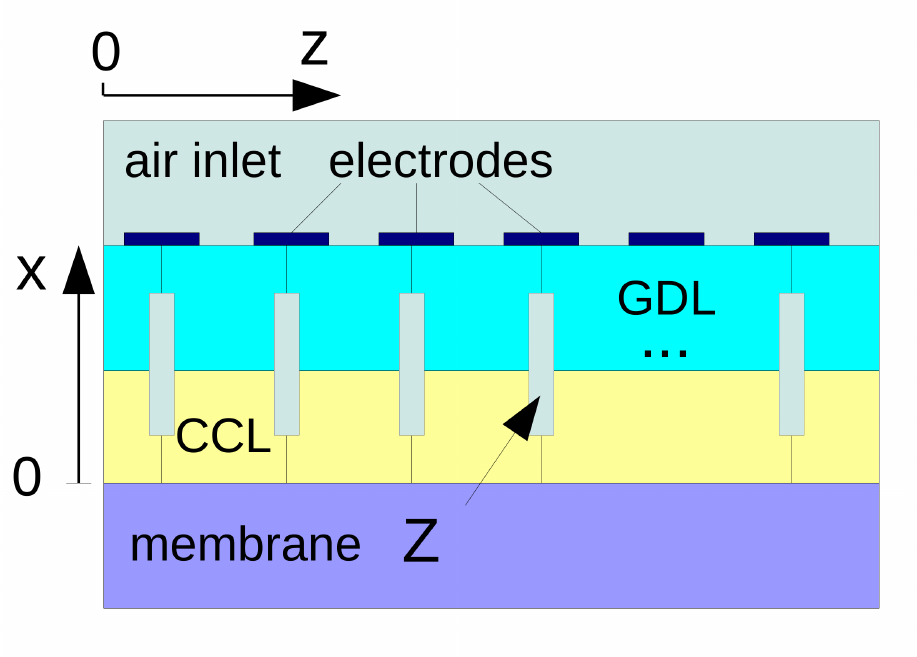}
		\caption{Schematic of the segmented cell cathode side with the straight channel
			and the system of coordinates. $Z$ stands for local impedance
			depending on the distance along the channel coordinate $z$.
		}
		\label{fig:sketch}
	\end{center}
\end{figure}

Schematic of the cell with the straight cathode channel is shown
in Figure~\ref{fig:sketch}.
Our main goal here is analysis of impedance of the GDL+channel oxygen transport system.
To simplify the model and to separate the GDL+channel impedance from impedance of oxygen transport
in the CCL, we will assume that the latter transport is fast.
In addition, for simplicity we will assume that the proton transport in the CCL is also fast.
The characteristic frequency of proton transport in the CCL is much higher
than the other frequencies in the system and hence this assumption does not
affect low-- and medium--frequency impedances which are of primary interest in this work.

\subsection{Proton charge conservation equation}
\label{sec:proton}

The proton charge conservation equation reads
\begin{equation}
   \Cdl\pdr{\eta}{t} + \pdr{j}{x} = - i_*\left(\dfrac{c}{\cref}\right)\lexp{\dfrac{\eta}{b}}
   \label{eq:etax}
\end{equation}
where
$\Cdl$ is the volumetric double layer capacitance (F~cm$^{-3}$),
$\eta$ is the positive by convention ORR overpotential,
$j$ is the local proton current density in the CCL,
$x$ is coordinate through the cell cathode counted from the membrane,
$i_*$ is the volumetric exchange current density (A~cm$^{-3}$),
$c$ is the local oxygen concentration,
$\cref$ is the reference oxygen concentration, and
$b$ is the ORR Tafel slope.

Since proton and oxygen transport in the CCL are fast, $\eta$ and $c$ are nearly independent of $x$.
Integrating Eq.\eqref{eq:etax} over $x$ from 0 to the CCL thickness $\lcat$ we come to
\begin{equation}
   \Cdl\lcat\pdr{\eta_0}{t} - j_0 = - i_*\lcat\left(\dfrac{c_1}{\cref}\right)\lexp{\dfrac{\eta_0}{b}}
   \label{eq:etax2}
\end{equation}
where
$\eta_0$ is the ORR overpotential at the membrane surface,
$j_0$ is the local cell current density, and
$c_1$ is the oxygen concentration at the CCL/GDL interface.

To simplify calculations, we introduce dimensionless variables
\begin{multline}
   \tit = \dfrac{t}{t_*}, \quad \tx = \dfrac{x}{\lcat},
                          \quad \tz = \dfrac{z}{L},
                          \quad \tc = \dfrac{c}{\cref},
                          \quad \tj = \dfrac{j}{i_*\lcat},\\
                               \teta = \dfrac{\eta}{b},
                          \quad \tD_b = \dfrac{4 F D_b \cref}{i_*\lcat^2},
                          \quad \tl_b = \dfrac{l_b}{\lcat},\\
                                \tom = \omega t_*,
                          \quad \tZ = \dfrac{Z i_* \lcat}{b}
   \label{eq:dless}
\end{multline}
where
\begin{equation}
   t_* = \dfrac{\Cdl b}{i_*},
   \label{eq:tast}
\end{equation}
is the characteristic time of double layer charging,
$z$ is the coordinate along the cathode channel,
$L$ is the channel length,
$D_b$ is the oxygen diffusion coefficient in the GDL,
$l_b$ is the GDL thickness,
$\omega$ is the angular frequency of the applied AC signal, and
$Z$ is the local impedance.

With the dimensionless variables \eqref{eq:dless}, Eq.\eqref{eq:etax2} takes the form
\begin{equation}
   \pdr{\teta_0}{\tit} - \tj_0 = - \tc_1 \expo
   \label{eq:tetax2}
\end{equation}
Substituting Fourier--transforms
\begin{equation}
   \begin{split}
      &\teta_0 = \teta^0 + \teta^1(\tom)\exp(\ri\tom\tit) \\
      &\tj_0 = \tj^0 + \tj^1(\tom)\exp(\ri\tom\tit)     \\
      &\tc_1 = \tc_1^0 + \tc_1^1(\tom)\exp(\ri\tom\tit)
   \end{split}
   \label{eq:Four}
\end{equation}
into Eq.\eqref{eq:tetax2},
expanding exponent in Taylor series, neglecting term with the perturbations product,
and subtracting the static equation, we get
equation relating the small perturbation amplitudes $\teta^1$, $\tj^1$ and $\tc_1^1$
in the $\tom$--space:
\begin{equation}
   \tj^1 = \ri\tom\teta^1 + \expo\left(\tc_1^1 + \tc_1^0\teta^1\right).
   \label{eq:teta1x}
\end{equation}
Here, the superscripts 0 and 1 mark the static variables and the small perturbation amplitudes,
respectively. Local cathode side impedance at a distance $\tz$ from the channel inlet
is given by
\begin{equation}
   \tZ = \dfrac{\teta^1}{\tj^1}
   \label{eq:tZdef}
\end{equation}
Dividing Eq.\eqref{eq:teta1x} by $\tj^1$, we obtain equation for $\tZ$:
\begin{equation}
   1 = \left(\ri\tom + \tc_1^0\expo\right)\tZ + \expo\dfrac{\tc_1^1}{\tj^1}
   \label{eq:tZ}
\end{equation}

Suppose that the perturbation of oxygen concentration is zero: $\tc_1^1=0$.
Physically, this is equivalent to fast oxygen transport in all transport medias
of the cathode side, including channel. Solution to Eq.\eqref{eq:tZ} is then
\begin{equation}
   \tZ_f = \dfrac{1}{\ri\tom + \tc_1^0\expo},
   \label{eq:tZ0_sol}
\end{equation}
which is impedance of a parallel $RC$--circuit, where the term $\ri\tom$ in denominator
describes contribution of the double layer capacitance and $\tc_1^0\expo$ is the
inverse faradaic resistivity of the catalyst layer.

The static oxygen concentration $\tc_1^0$ is related to the channel concentration
$\tc_h^0$ as
\begin{equation}
   \tc_1^0 = \tc_h^0\left(1 - \dfrac{\tj^0}{\tjlim}\right)
   \label{eq:tc10}
\end{equation}
where
\begin{equation}
   \jlim = \dfrac{4 F D_b c_h^0}{l_b}
   \label{eq:jlim}
\end{equation}
is the local limiting current density due to oxygen transport in the GDL.
Eq.\eqref{eq:tc10} is solution of the static version of equation for oxygen transport
in the GDL, see Eq.\eqref{eq:cbx} below.
The concentration $\tc_h^0$ varies along the channel; numerical model\cite{Kulikovsky_19a}
shows that unless the mean cell current density is small, this variation
to a good approximation is linear:
\begin{equation}
   \tc_h^0 \simeq 1 - \dfrac{\tz}{\lam}
   \label{eq:tch0z}
\end{equation}
Note that Eq.\eqref{eq:tch0z} is an approximation
\footnote{At low cell currents, the shape of local oxygen concentration
	is exponential\cite{Kulikovsky_03b}:
	$$
	\tc_h^0 = \left(1 - \dfrac{1}{\lam}\right)^{\tz}
	$$
	The model can be reformulated using the above equation for $\tc_h^0$.
	However, this equation is valid at the low cell currents only, while
	Eq.\eqref{eq:tch0z} works better at higher cell currents, which are of practical interest.
}
the exact shape of $\tc_h^0(\tz)$
should be calculated using a numerical model\cite{Chevalier_18b,Kulikovsky_19a}.

To simplify calculations, we will ignore the factor $1 - {\tj^0}/{\tjlim}$ in Eq.\eqref{eq:tc10},
assuming that $\jlim$ is large and the variation of static oxygen concentration
across the GDL is negligible.
With this, Eqs.\eqref{eq:teta1x}, \eqref{eq:tZ} transform to
\begin{equation}
   \tj^1 = \ri\tom\teta^1 + \expo\left(\tc_1^1 + \left(1 - \dfrac{\tz}{\lam}\right)\teta^1\right)
   \label{eq:teta1xb}
\end{equation}
\begin{equation}
    1 = \left(\ri\tom + \expo\left(1 - \dfrac{\tz}{\lam}\right)\right)\tZ + \expo\dfrac{\tc_1^1}{\tj^1}
	\label{eq:tZb}
\end{equation}
As discussed above, the non--faradaic oxygen transport contributions to local
impedance gives the term with $\tc_1^1$.
In order to calculate $\tc_1^1$, we need to consecutively solve equations for oxygen
transport in the GDL and channel, as discussed in the next section.

For further references we need a total faradaic impedance $\tZ_{f,tot}$ of the cathode,
taking into account variation of oxygen concentration along the channel.
Suppose that the cell is divided into $N\to\infty$ virtual
segments. Local current in each segment flows in the through--plane direction, hence
local faradaic impedances are connected in parallel. Thus, $\tZ_{f,tot}$ is given by
\begin{equation}
   \tZ_{f,tot} = \left(\int_0^1\dfrac{d\tz}{\tZ_f}\right)^{-1}
   \label{eq:tZftot_def}
\end{equation}
Electron conductivity of the cell components is assumed to be large and hence $\teta^0$
is independent of the coordinate $\tz$.
Substituting Eq.\eqref{eq:tch0z} into Eq.\eqref{eq:tZ0_sol}, we get the local faradaic
impedance $\tZ_f(\tz)$. Calculating integral, from Eq.\eqref{eq:tZftot_def} we find
\begin{equation}
   \tZ_{f,tot} = \dfrac{1}{\ri\tom + \left(1 - \dfrac{1}{2\lam}\right)\expo}
   \label{eq:tZftot}
\end{equation}

\subsection{Oxygen transport in the GDL}

Oxygen transport in the gas--diffusion layer is described by the diffusion equation
\begin{multline}
   \pdr{c_b}{t} - D_b\pddr{c_b}{x} = 0, \quad
   \left.D_b\pdr{c_b}{x}\right|_{x=\lcat+} = \dfrac{j_0}{4F},  \\
    c_b(\lcat + l_b) = c_h,
   \label{eq:cbx}
\end{multline}
where
$c_b$ is the oxygen concentration in the GDL and
$c_h$ is the oxygen concentration in channel.
The left boundary condition for Eq.\eqref{eq:cbx} means that
the oxygen flux on the GDL side of the CCL/GDL interface equals the local current density
in the cell. This condition agrees with the assumption of fast oxygen transport
in the CCL.

With the dimensionless variables Eq.\eqref{eq:dless}, Eq.\eqref{eq:cbx} takes the form
\begin{multline}
   \mu^2\pdr{\tc_b}{\tit} - \tD_b\pddr{\tc_b}{\tx} = 0,
   \quad \left.\tD_b\pdr{\tc_b}{\tx}\right|_{\tx=1+} = \tj_0, \\
    \tc_b(1 + \tl_b) = \tc_h,
   \label{eq:tcbx}
\end{multline}
where the dimensionless parameter $\mu$ is given by
\begin{equation}
   \mu = \sqrt{\dfrac{4F\cref}{\Cdl b}}.
     \label{eq:mu}
\end{equation}

Eq.\eqref{eq:tcbx} is linear and hence the equation for the small perturbation
amplitude $\tc_b^1$ is
\begin{multline}
   \tD_b\pddr{\tc_b^1}{\tx} = \ri\tom\mu^2\tc_b^1,
   \quad \left.\tD_b\pdr{\tc_b^1}{\tx}\right|_{\tx=1+} = \tj^1, \\
    \tc_b^1\left(1 + \tl_b\right) = \tc_h^1
   \label{eq:tcb1x}
\end{multline}
where $\tc_h^1$ is the oxygen perturbation in channel (see below).

Solution to Eq.\eqref{eq:tcb1x} is
\begin{multline}
   \tc_b^1 =  - \dfrac{\tj^1\sinh\left( \mu \sqrt{\sqa}\,\left(1 + \tl_b - \tx\right)\right)}
                      {\mu\sqrt{\ri\tom\tD_b}\cosh\left(\mu\tl_b\sqrt{\sqa}\right)} \\
              + \dfrac{\tc_h^1\cosh\left(\, \mu \sqrt{\sqa} (1 - \tx)\right)}
                      {\cosh\left(\mu\tl_b\sqrt{\sqa}\right)}
   \label{eq:tcb1xsol}
\end{multline}
Setting here $\tx=1$, we get the perturbation amplitude of oxygen concentration
at the CCL/GDL interface $\tc_b^1(1) = \tc_1^1$
\begin{equation}
   \tc_1^1 =  - \dfrac{\tj^1\tanh\left(\phi\right)}
                      {\psi}
              + \dfrac{\tc_h^1}
                      {\cosh\left(\phi\right)}
   \label{eq:tcb1sol}
\end{equation}
where $\phi$ and $\psi$ are auxiliary dimensonless parameters
\begin{equation}
   \begin{split}
      &\phi = \mu\tl_b\sqrt{\sqa} \\
      &\psi = \mu\sqrt{\ri\tom\tD_b}
   \end{split}
   \label{eq:phipsi}
\end{equation}

\subsection{Oxygen transport in channel}

To a good approximation, oxygen mass transport in the channel can be
described by the 1d + 1d plug flow equation:
\begin{equation}
   \pdr{c_h}{t} + v\pdr{c_h}{z} = - \left.\dfrac{D_b}{h}\pdr{c_b}{x}\right|_{x=\lcat + l_b}
   \label{eq:chz}
\end{equation}
where
$h$ is the channel depth.
The right side of Eq.\eqref{eq:chz} is the oxygen diffusive flux
in the GDL at the channel/GDL interface representing oxygen ``sink'' from the channel.

With the dimensionless variables \eqref{eq:dless}, Eq.\eqref{eq:chz} reads
\begin{equation}
   \xi^2\pdr{\tc_h}{\tit} + \lam\tJ\pdr{\tc_h}{\tz} = - \left.\tD_b\pdr{\tc_b}{\tx}\right|_{\tx=1 + \tl_b}
   \label{eq:tchz}
\end{equation}
where
$\tJ$ is the mean current density in the cell
\begin{equation}
   \tJ = \int_0^1\tj_0 d\tz,
   \label{eq:tJ}
\end{equation}
$\xi$ is the dimensionless parameter, and $\lam$ is the stoichiometry of air flow
\begin{equation}
    \xi = \sqrt{\dfrac{4 F h \cref}{\Cdl \lcat b}},\quad \lam = \dfrac{4 F h v \cref}{L J}.
    \label{eq:xilam}
\end{equation}

Eq.\eqref{eq:tchz} is linear and we can immediately write down equation for the perturbation
amplitude $\tc_h^1$:
\begin{equation}
   \lam\tJ \pdr{\tc_h^1}{\tz} =  - \ri\tom\xi^2\tc_h^1 - \left.\tD_b\pdr{\tc_b^1}{\tx}\right|_{\tx=1 + \tl_b}
   \label{eq:tch1z}
\end{equation}
Differentiating Eq.\eqref{eq:tcb1xsol}, we find the flux $\tD_b\pdra{\tc_b^1}{\tx}|_{\tx=1 + \tl_b}$
and Eq.\eqref{eq:tch1z} takes the form
\begin{multline}
   \lam\tJ \pdr{\tc_h^1}{\tz} =  - \ri\tom\xi^2\tc_h^1
                                 - \tj^1 \cosh\left(\phi\right)^{-1}
                                 - \tc_h^1 \psi\,\tanh\left(\phi\right), \\
    \tc_h^1(0) = 0
   \label{eq:tch1z2}
\end{multline}
where $\tj^1$ is a function of coordinate $\tz$ and of $\tc_h^1$. To find
the explicit dependence $\tj^1(\tz)$ we substitute \eqref{eq:tcb1sol} into Eq.\eqref{eq:teta1xb};
solving the resulting equation for $\tj^1$ we get
\begin{multline}
   \tj^1 = \left(\dfrac{\expo\tanh\left(\phi\right)}{\psi} + 1 \right)^{-1} \\
              \times\left(\left(\ri\tom + \expo\left(1 - \dfrac{\tz}{\lam}\right)\right) \teta^1
            + \dfrac{\expo\tc_h^1}{\cosh(\phi)}\right)
   \label{eq:tj01_sol}
\end{multline}

\subsection{Solution procedure and total cathode impedance}

Substituting $\tj^1$, Eq.\eqref{eq:tj01_sol}, into Eq.\eqref{eq:tch1z2} and solving
the resulting equation we find the amplitude of oxygen concentration perturbation along
the channel
\begin{equation}
   \tc_h^1(\tz) = \dfrac{\teta^1}{B}\left(\left(\dfrac{A\lam\tJ}{B} + C\right)\left(\lexp{\dfrac{B\tz}{\lam\tJ}} - 1\right)
                 - A\tz\right)
    \label{eq:tch1sol}
\end{equation}
where the independent of $\tz$ coefficients $A$, $B$ and $C$ are given by
\begin{equation}
   A = \dfrac{\psi\expo}{\lam\cosh(\phi)\left(\psi + \expo\tanh(\phi)\right)}
   \label{eq:A}
\end{equation}
\begin{equation}
   B = - \ri\tom\xi^2 - \psi\tanh(\phi) - \dfrac{\lam A}{\cosh(\phi)}
   \label{eq:B}
\end{equation}
\begin{equation}
   C = -\dfrac{\psi\left(\ri\tom + \expo\right)}
              {\cosh(\phi)\left(\psi + \expo\tanh(\phi)\right)}
   \label{eq:C}
\end{equation}
Setting in Eq.\eqref{eq:tch1sol} $\teta^1 = \tZ\tj^1$ and using the result in Eq.\eqref{eq:tcb1sol}
we get a formula for $\tc_1^1$, which linearly depends on $\tZ$. Substituting this $\tc_1^1$
into Eq.\eqref{eq:tZb}, we obtain a linear algebraic equation for local impedance $\tZ$.
Solving this equation, we find local impedance, which includes faradaic and oxygen
transport in the channel and GDL processes
\begin{equation}
   \tZ(\tz) = \dfrac{B^2}{D_{loc}}\left(\cosh(\phi) + \dfrac{\expo\sinh(\phi)}{\psi}\right)
   \label{eq:tZloc}
\end{equation}
where
\begin{multline}
   D_{loc} = \expo \left(\lam\tJ A + B C \right)\left(\lexp{\dfrac{B\tz}{\lam\tJ}} - 1 \right) \\
               + B^2\left(\ri\tom + \expo\left(1 - \dfrac{\tz}{\lam}\right)\right)\cosh(\phi)
               - \expo B A\tz.
   \label{eq:Dloc}
\end{multline}
Note that $D_{loc}$ contains complex exponent $\lexp{{B\tz}/({\lam\tJ})}$ leading
to oscillations of $\tZ$ along the channel coordinate $\tz$ (Ref.\cite{Kulikovsky_15g}).

The total cathode side impedance is given by
\begin{equation}
   \tZ_{tot} = \left(\int_0^1\dfrac{d\tz}{\tZ}\right)^{-1}
   \label{eq:tZtot}
\end{equation}
Calculation of integral leads to
\begin{equation}
   \tZ_{tot} = \dfrac{\lam B^3}{D_{tot}}\left(\cosh(\phi) + \dfrac{\expo\sinh(\phi)}{\psi}\right)
   \label{eq:tZtotal}
\end{equation}
where
\begin{multline}
   D_{tot} = \lam^2\tJ\expo\left(\lam\tJ A + B C\right)\left(\lexp{\dfrac{B}{\lam\tJ}} - 1 \right) \\
                    + \left((\ri\tom + \expo)\lam - \expo/2\right) B^3\cosh(\phi) \\
              - \lam\expo B \left(\lam\tJ A + B(A/2 + C)\right).
   \label{eq:Dtot}
\end{multline}
The closed form of integral in Eq.\eqref{eq:tZtot} is a key point leading to analytical
formula for $\tZ_{tot}$. At high frequencies of the AC signal, $\tc_h^1$ and local impedance $\tZ$
are rapidly oscillating functions of $\tz$, which requires a lot of steps in
numerical solution of Eq.\eqref{eq:tch1z} and makes it difficult
numerical calculation of Eq.\eqref{eq:tZtot}. Analytical result
for integral in Eq.\eqref{eq:tZtot} solves the problem.

\section{Results and Discussion}

Equations of the previous Section contain mean current density in the cell $\tJ$
and the total static potential loss (overpotential) $\teta^0$. These parameters
are related by the polarization curve, which could be obtained from
solution of the static version of equations \eqref{eq:etax}, \eqref{eq:cbx} and \eqref{eq:chz}.
The solution is Tafel--like equation corrected for
the finite flow stoichiometry $\lam$ (Ref.\cite{Kulikovsky_03d}):
\begin{equation}
   - \lam\lnl{1-\dfrac{1}{\lam}}\tJ = \expo
   \label{eq:vcc}
\end{equation}
Note that Eq.\eqref{eq:vcc} does not include the effect of cell ohmic resistivity.
If experimental polarization curve is available, an IR--corrected numerical relation
between $\teta^0$ and $\tJ$ could be used instead of Eq.\eqref{eq:vcc}.
A more accurate numerical approximation for the polarization curve could be
obtained using the model\cite{Chevalier_18b,Kulikovsky_19a}.

\begin{figure}
\begin{center}
\includegraphics[scale=0.5]{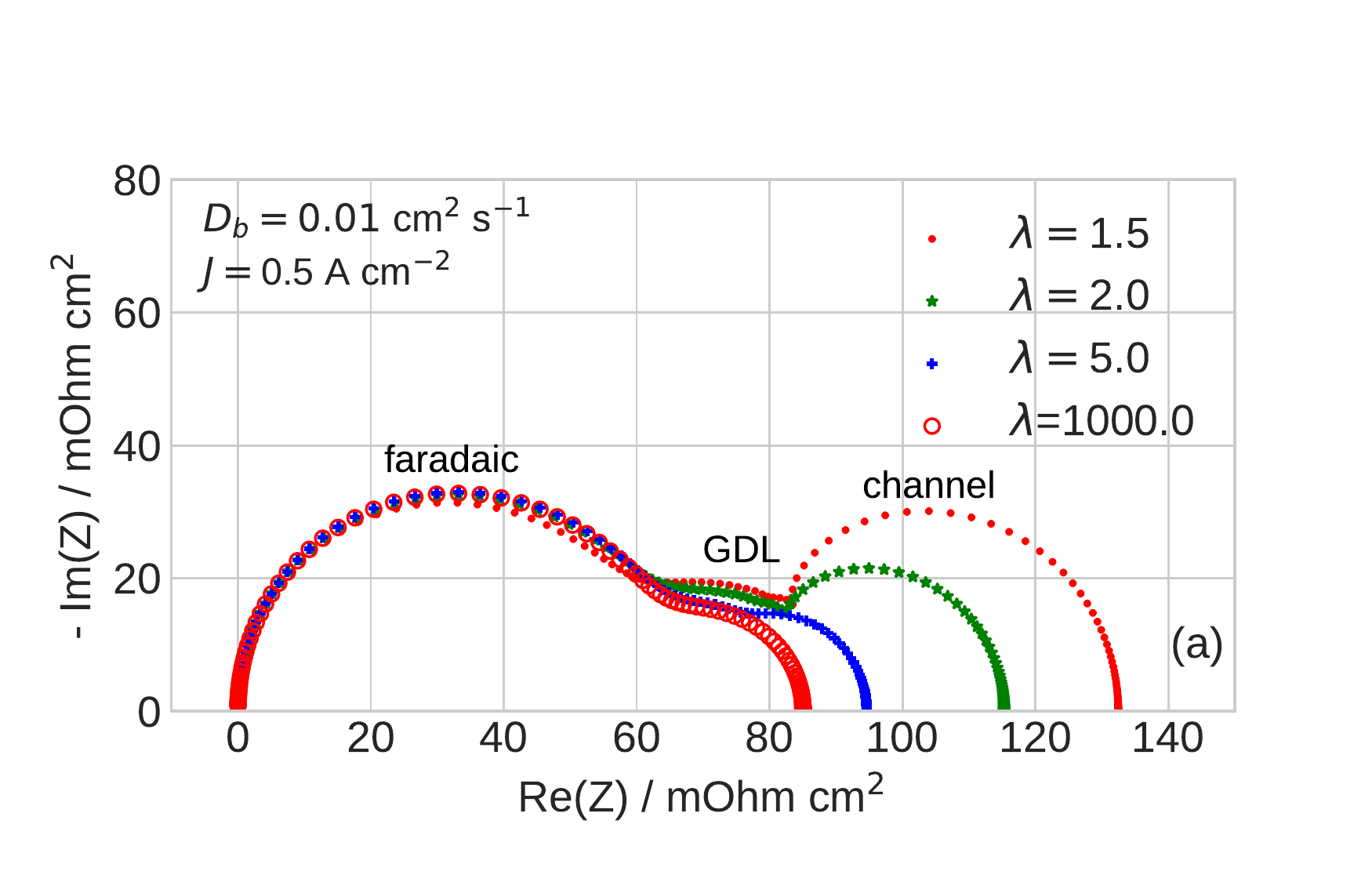}
\includegraphics[scale=0.5]{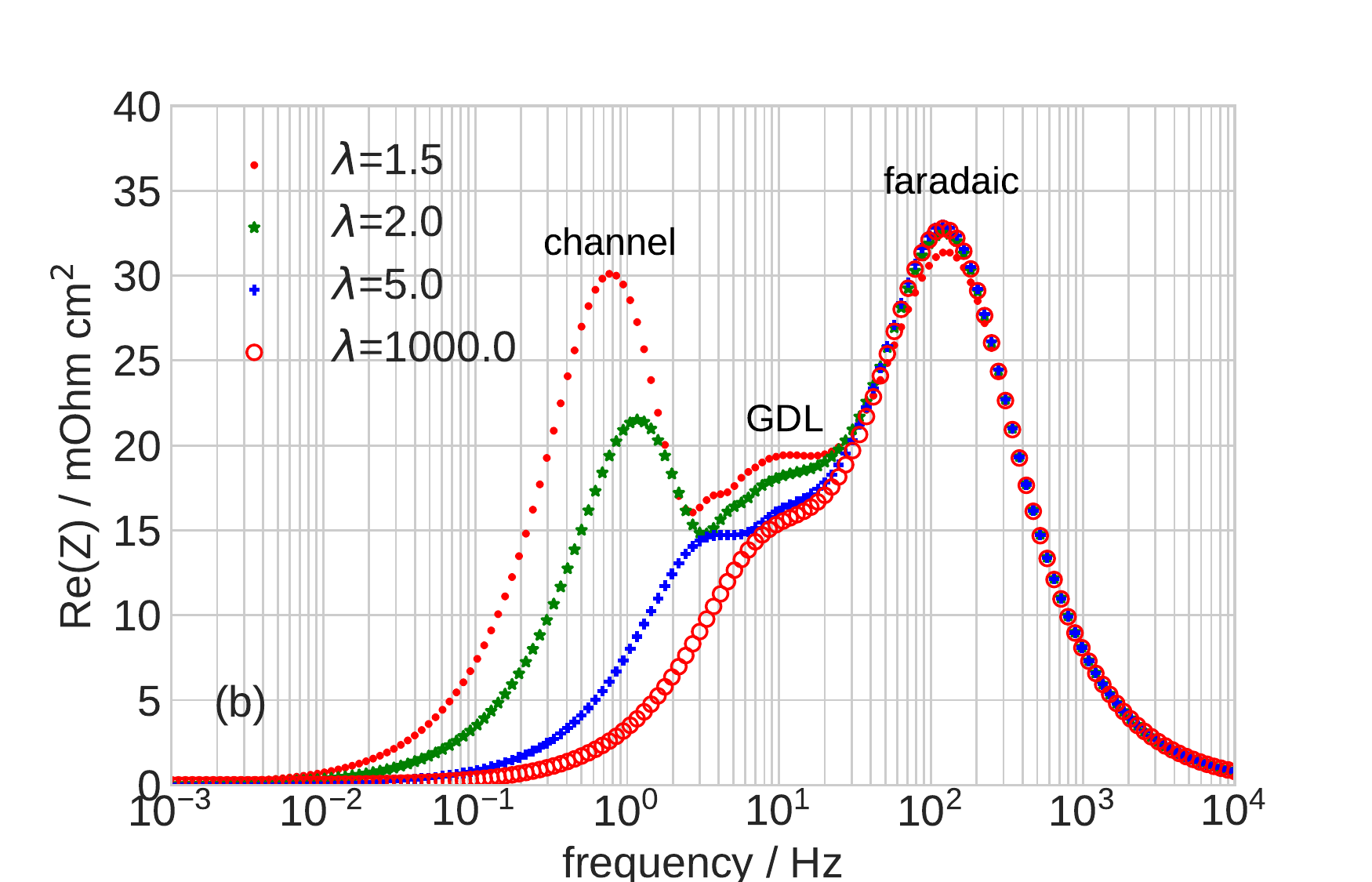}
\caption{(a) Nyquist spectra of the total cathode side impedance, Eq.\eqref{eq:tZtotal},
   for the indicated air flow stoichiometries $\lam$. The cell current density
   is 0.5~A~cm$^{-2}$.
   (b) Frequency dependence of imaginary part of impedance in (a).
  }
\label{fig:Nyqfull}
\end{center}
\end{figure}

Eq.\eqref{eq:vcc} allows us to eliminate $\expo$ from Eq.\eqref{eq:tZtotal}.
Nyquist spectra of total cathode side impedance $Z_{tot}$ for the parameters
listed in Table~\ref{tab:parms}, the mean cell current
density $J=0.5$~A~cm$^{-2}$ and several stoichiometries of the air flow
are shown in Figure~\ref{fig:Nyqfull}a.
Figure~\ref{fig:Nyqfull}b shows the frequency dependence of $\Im{Z_{tot}}$.
As can be seen, the impedance consists of
three arcs, of which the low--frequency (LF) one strongly depends on $\lam$
(Figure~\ref{fig:Nyqfull}a).
In the pioneering experiments of Schneider \etal this arc has
been associated with the oxygen transport in channel\cite{Ingo_07a,Ingo_07b}.
The LF arc vanishes as $\lam\to\infty$ (Figures~\ref{fig:Nyqfull}a,b).

\begin{figure}
\begin{center}
\includegraphics[scale=0.5]{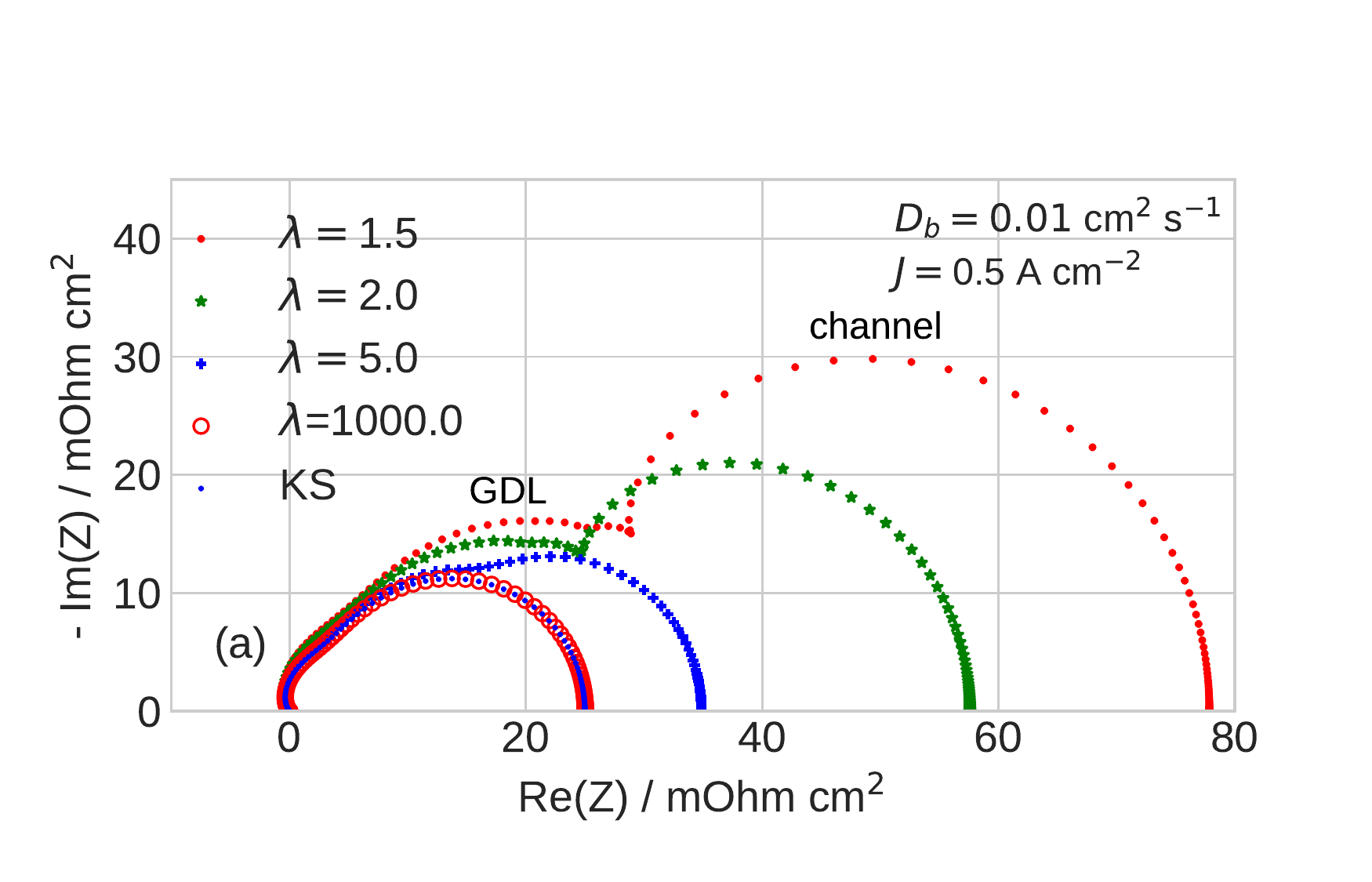}
\includegraphics[scale=0.5]{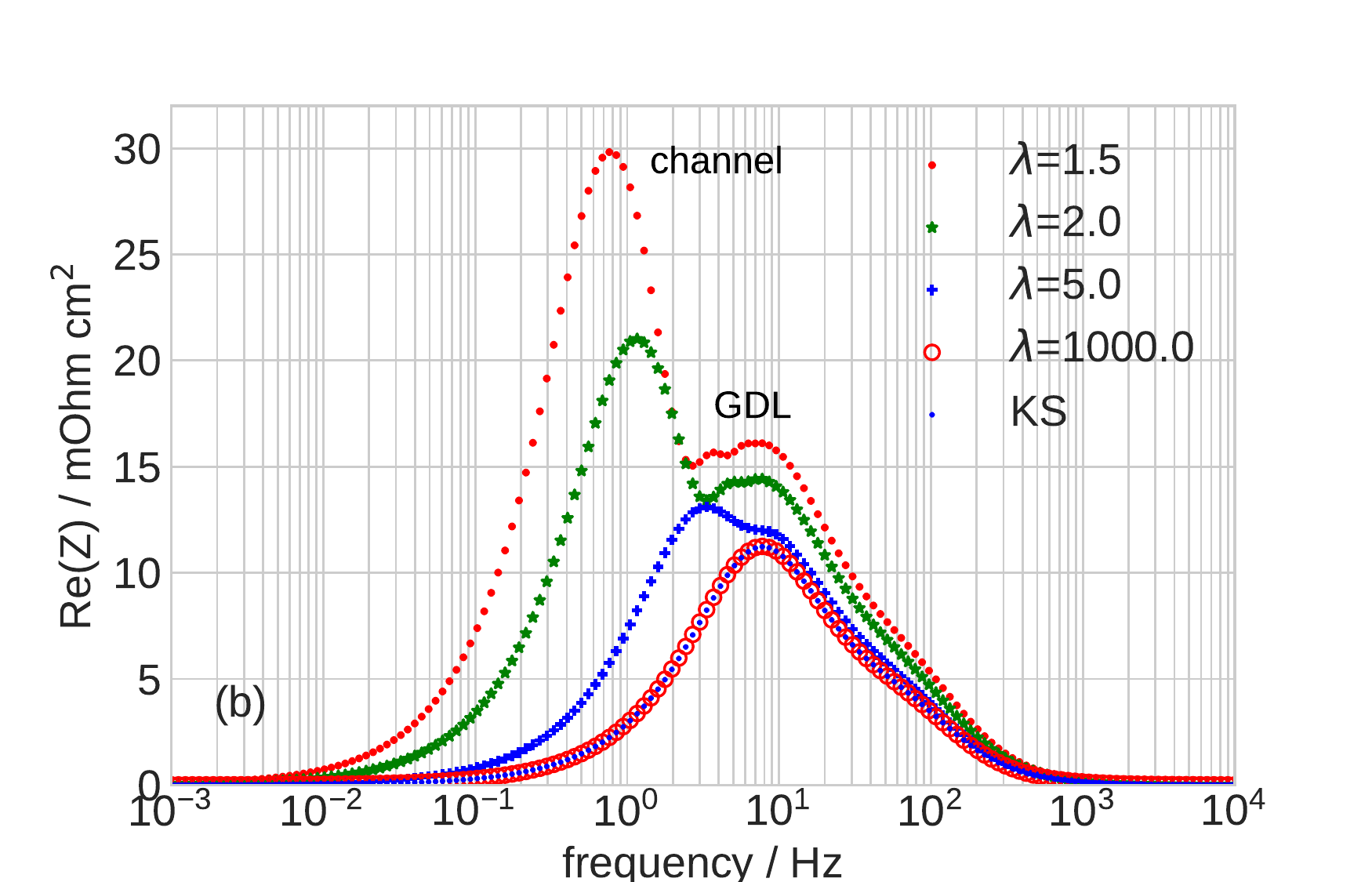}
\caption{(a) Nyquist spectra of the GDL+channel impedance, Eq.\eqref{eq:tZgdlc},
   for the indicated air flow stoichiometries $\lam$.
   The cell current density is $J=0.5$~A~cm$^{-2}$. KS marks the spectrum of
   Eq.\eqref{eq:tZgdl}.
   (b) Frequency dependence of imaginary part of impedance in (a).
  }
\label{fig:NyqBode}
\end{center}
\end{figure}

The medium--frequency (MF) arc in Figure~\ref{fig:Nyqfull}a
represents oxygen transport in the GDL.
This arc is not fully seen  due to masking effect of
the third, high--frequency (HF) faradaic arc (Figure~\ref{fig:Nyqfull}).
To emphasize the GDL arc, we subtract  the total faradaic impedance \eqref{eq:tZftot}
from the total cathode impedance, Eq.\eqref{eq:tZtotal}. This leads to the
GDL+channel impedance $\tZ_{gdlc}$:
\begin{equation}
   \tZ_{gdlc} = \tZ_{tot} - \dfrac{1}{\ri\tom + \left(1 - \dfrac{1}{2\lam}\right)\expo}
   \label{eq:tZgdlc}
\end{equation}
Nyquist spectra of Eq.\eqref{eq:tZgdlc} for the same parameters
exhibit two arcs;
now the left, GDL arc is fully resolved  (Figure~\ref{fig:NyqBode}a).
Frequency dependence of $\Im{Z_{gdlc}}$
is shown in Figure~\ref{fig:NyqBode}b. Two peaks in Figure~\ref{fig:NyqBode}b correspond to
the channel and GDL arcs in Figure~\ref{fig:NyqBode}a.
At low stochiometry of the air flow,
the GDL arc is located at the right ``wing'' of the channel arc; the latter
contributes to imaginary part of the GDL impedance (Figure~\ref{fig:NyqBode}b).
However, as $\lam$ increases, the channel arc vanishes and only GDL arc is left
(the curves for $\lam=1000$ in Figures~\ref{fig:NyqBode}a,b).

\begin{table}
\small
\begin{tabular}{|l|c|}
\hline
   GDL thickness $l_b$, cm              & 0.023  \\
   Catalyst layer thickness $\lcat$, cm & $10\cdot 10^{-4}$ (10 $\mu$m)       \\
   ORR Tafel slope $b$, mV              &  30 \\
   Double layer capacitance $\Cdl$, F~cm$^{-3}$  &  20   \\
   GDL oxygen diffusivity $D_b$, cm$^2$~s$^{-1}$ &  0.01 \\
   Cell current density $j_0$, A~cm$^{-2}$  &  0.1   \\
   Pressure                             & Standard   \\
   Cell temperature $T$, K              & 273 + 80   \\
   Aur flow stoichiometry $\lam$        & 2.0 \\
\hline
\end{tabular}
\caption{The base--case cell parameters used in calculations.
  }
\label{tab:parms}
\end{table}
%


Equation for GDL impedance in the limit of infinite air flow stoichiometry
has been derived by Kulikovsky and Shamardina\cite{Kulikovsky_15g}.
In the notations of this work, Eq.(36) of Ref.\cite{Kulikovsky_15g} is
\begin{equation}
   \tZ_{gdl} = \dfrac{\tanh\left(\mu\tl_b\sqrt{\sqa}\right)}
                     {\mu \sqrt{\ri\tom\tD_b}\left(1 - \tJ\tl_b/\tD_b\right)
                      \left(1 + \ri\tom/\tJ\right)}
   \label{eq:tZgdl0}
\end{equation}
The frequency--independent factor $(1 - \tJ\tl_b/\tD_b)$ in denominator describes the growth
of static resistivity upon approaching the limiting current density
$\tD_b/\tl_b$ due to oxygen transport in the GDL.
In this work, we assume that the limiting current is large and the factor $(1 - \tJ\tl_b/\tD_b)$ can
be replaced by unity. With this, Eq.\eqref{eq:tZgdl0} simplifies to
\begin{equation}
   \tZ_{gdl} = \dfrac{\tanh\left(\mu\tl_b\sqrt{\sqa}\right)}
                     {\mu \sqrt{\ri\tom\tD_b}\left(1 + \ri\tom/\tJ\right)}
   \label{eq:tZgdl}
\end{equation}
Eq.\eqref{eq:tZgdl} can be directly obtained from Eq.\eqref{eq:tZgdlc}
by passing to the limit $\lam \to \infty$.
The spectrum of Eq.\eqref{eq:tZgdl} (KS--spectrum) is shown in Figures~\ref{fig:NyqBode}a,b
by small blue dots. As expected, the spectrum of Eq.\eqref{eq:tZgdlc} for $\lam=1000$
(red open circles, Figure~\ref{fig:NyqBode}a,b) is practically
indistinguishable with the spectrum of Eq.\eqref{eq:tZgdl}.

Of particular interest is the ``non--Warburg'' factor $(1 + \ri\tom/\tJ)$
in denominator of Eq.\eqref{eq:tZgdl}.
Without this factor, Eq.\eqref{eq:tZgdl} is equivalent
to the Warburg finite--length impedance\cite{Lasia_book_14}.
``Non--Warburg'' shape of the GDL Nyquist arc (HF part
of this arc looks like an elephant's trunk, Figure~\ref{fig:NyqBode}a)
is due to the effect of charging double layer capacitance
in the CCL\cite{Kulikovsky_15g}. In other words, the GDL impedance ``feels''
the double layer capacitance of the attached CCL, since CCL provides a boundary
condition for oxygen transport through the GDL. Similar capacitive correction
to the classic Warburg impedance has been discussed
by Barbero\cite{Barbero_16} in the context of Poisson--Nernst--Plank model
for the planar electrode. It is worth noting a detailed study of the effect
of non--capacitive boundary conditions on Warburg impedance published
recently by Huang\cite{Huang_18}.

\begin{figure}
\begin{center}
\includegraphics[scale=0.5]{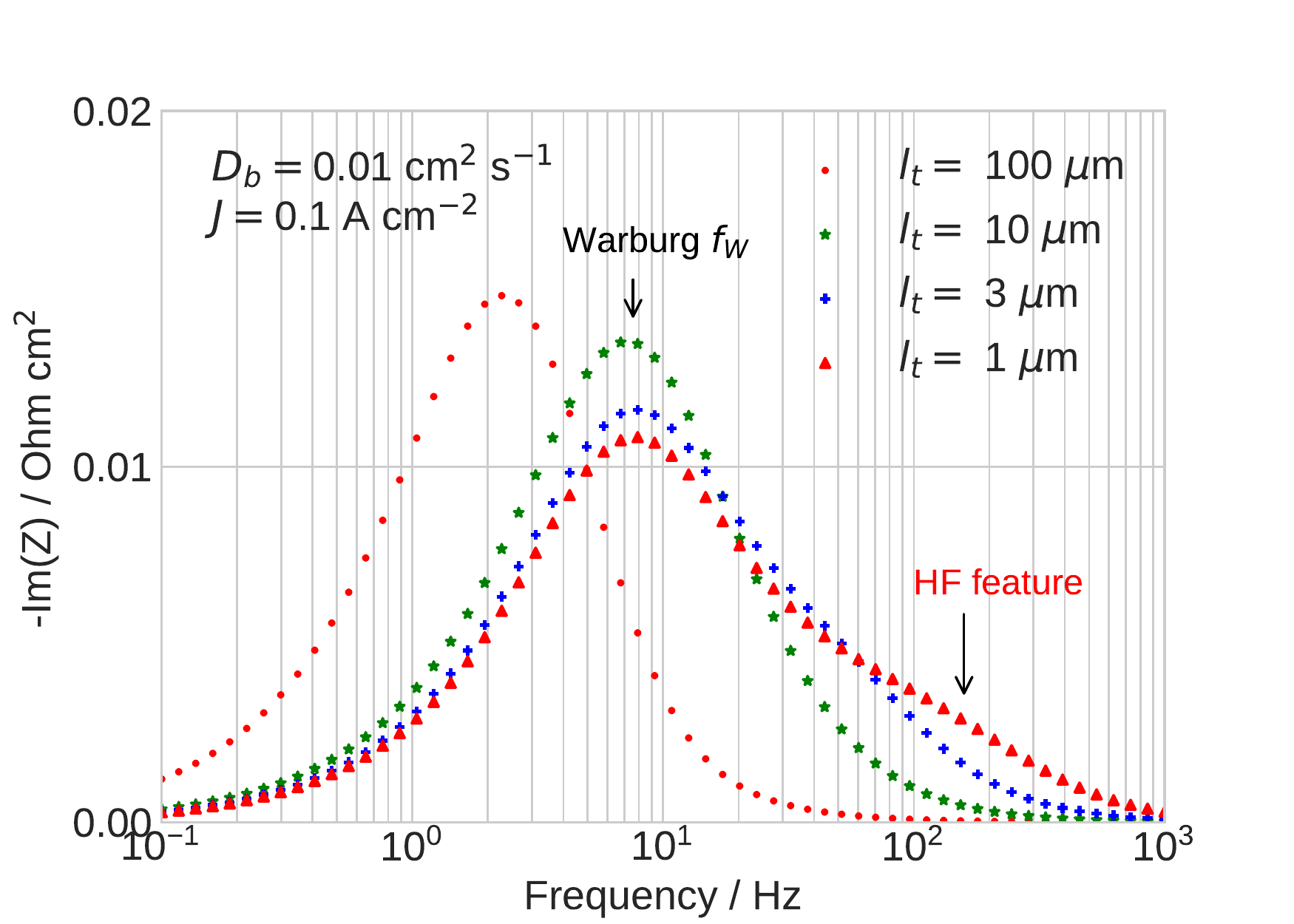}
\caption{Frequency dependence of imaginary part of the GDL impedance,
   Eq.\eqref{eq:tZgdl}, for the indicated values of CCL thickness $\lcat$.
   The cell current density is $0.1$~A~cm$^{-2}$. The volumetric
   double layer capacitance $\Cdl$ is assumed to be constant for
   all the curves.
  }
\label{fig:flcat}
\end{center}
\end{figure}

The non--Warburg factor $(1 + \ri\tom/\tJ)$
in Eq.\eqref{eq:tZgdl} leads to important effects.
In the dimension form, $\ri\tom/\tJ = \ri\omega\Cdl\lcat b/J$,
i.e., this factor includes the superficial double layer capacitance $\Cdl\lcat$ (F~cm$^{-2}$)
of the CCL. Under constant volumetric DL capacitance $\Cdl$ (F~cm$^{-3}$),
this results in dependence
of the GDL spectrum on the catalyst layer thickness $\lcat$.
Figure~\ref{fig:flcat} shows that with typical for Pt/C--based PEMFCs CCL thickness
in the range of 10 $\mu$m to 1 $\mu$m, the effect of non-Warburg factor on the
characteristic frequency $f_{gdl}$ of GDL spectrum is small and $f_{gdl}$
is close to the Warburg finite--length frequency $f_W$:
\begin{equation}
   f_W = \dfrac{2.54 D_b}{2\pi l_b^2}
   \label{eq:fW}
\end{equation}
However, for CCL thickness on the order of 100 $\mu$m typical for cells
with non--Pt catalysts\cite{Kulikovsky_20a}, the non--Warburg factor strongly shifts the
frequency $f_{gdl}$ to lower values (Figure~\ref{fig:flcat}).
This may lead to partial overlapping
of the GDL and channel peaks of the respective imaginary parts
(cf. Figure~\ref{fig:NyqBode}b, the curve for $\lam=5$).
Figure~\ref{fig:flcat} also shows that with the decrease in CCL thickness,
the width of the GDL peak increases. Moreover, at low $\lcat$, the non--Warburg
factor leads to formation of a distinct high--frequency feature
in the spectrum (Figure~\ref{fig:flcat}).
The effects in Figure~\ref{fig:flcat} are particularly important for identification
of DRT peaks in the high-- and low--Pt PEM fuel cell spectra. The thickness of a low--Pt CCL
is typically three to four times less than the high--Pt CCL thickness, while the volumetric
$\Cdl$ in both types of cells is the same.

\begin{figure}
\begin{center}
\includegraphics[scale=0.42]{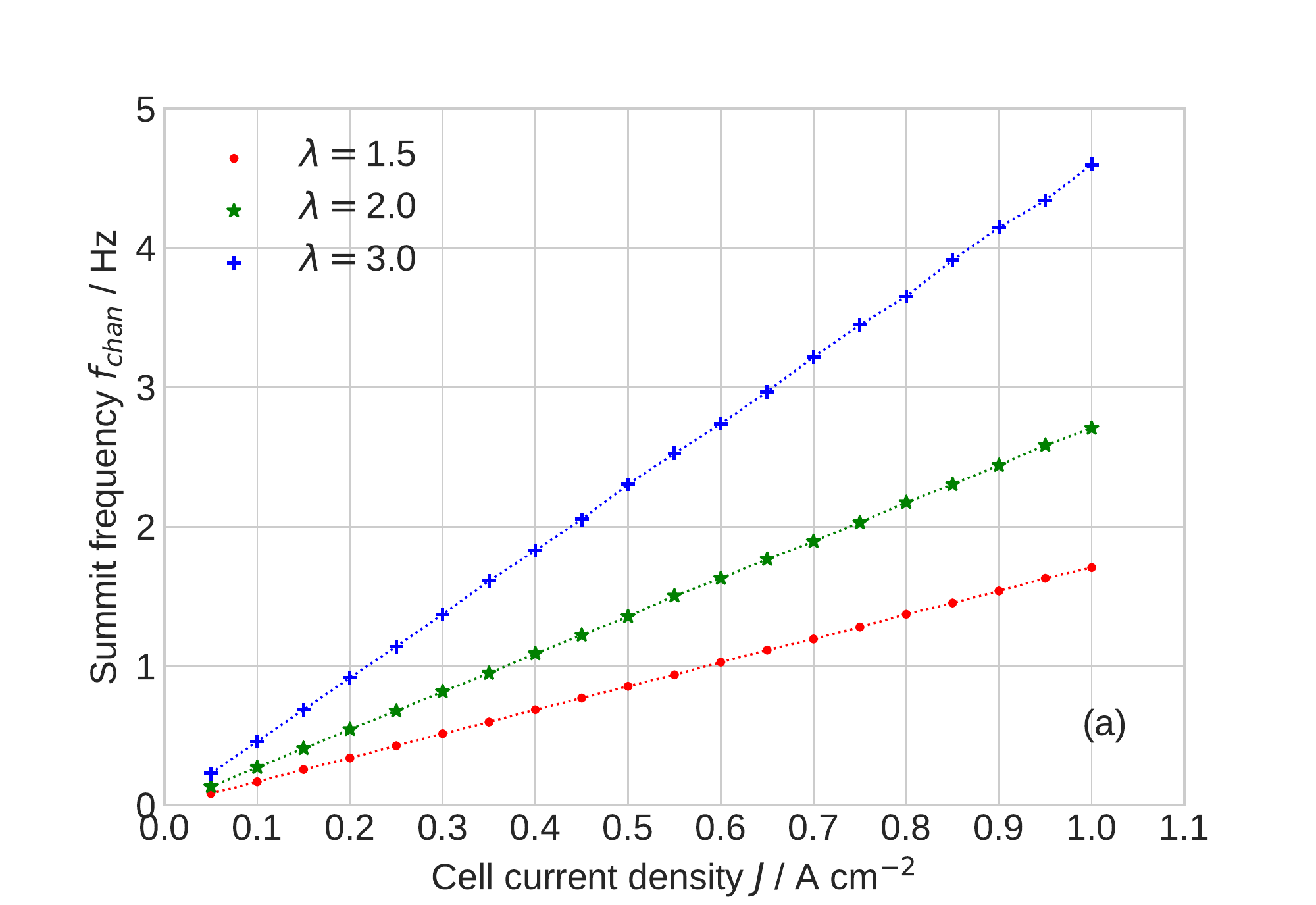}
\includegraphics[scale=0.42]{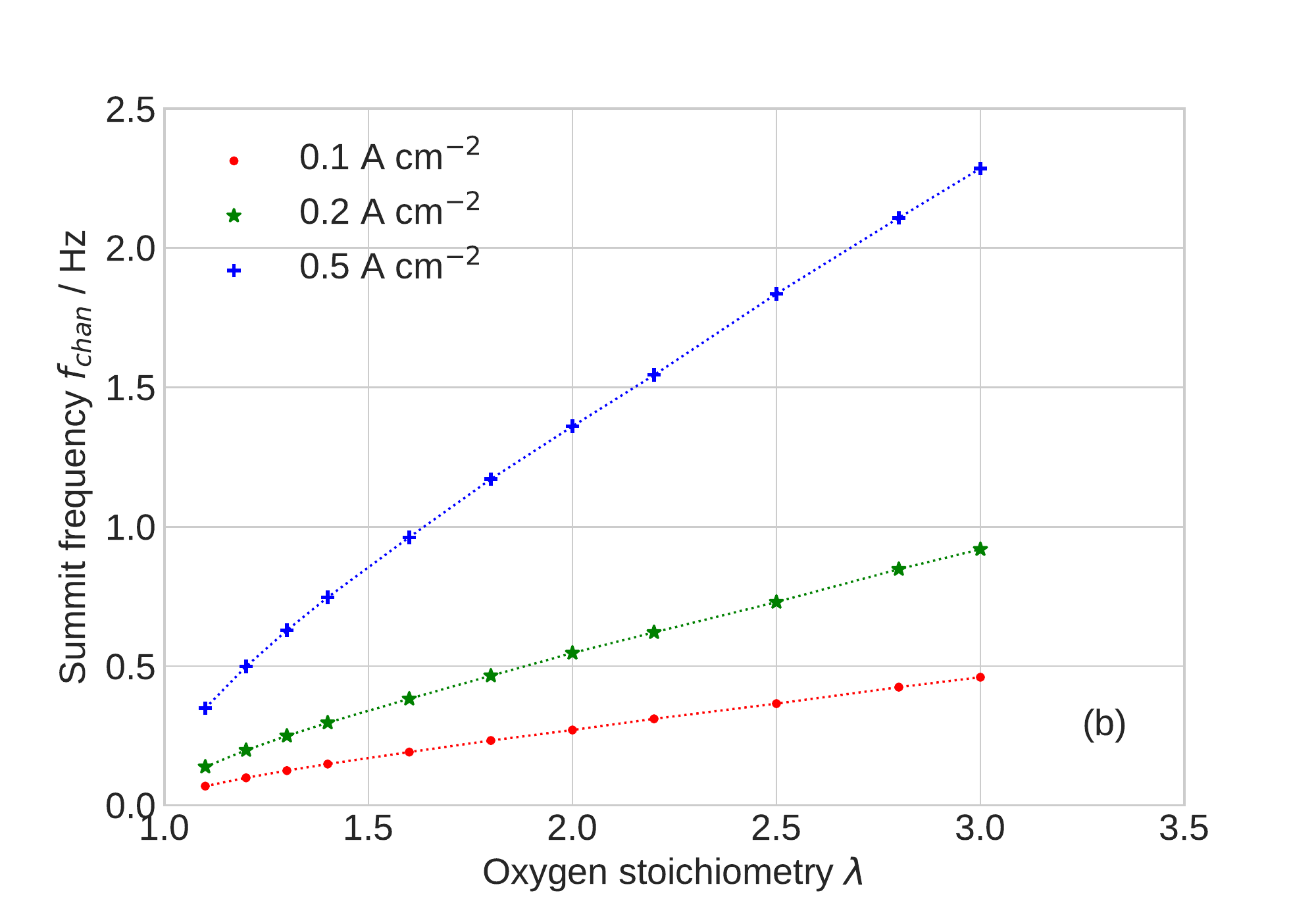}
\caption{(a) The characteristic frequency of channel impedance $f_{chan}$
   vs cell current
   density for the indicated air flow stoichiometries.
   (b) The frequency $f_{chan}$ vs air flow stoichiometry for the indicated
   cell current densities.
}
\label{fig:fchan}
\end{center}
\end{figure}

Finally, a formula for ``pure'' channel impedance $\tZ_{chan}$ can be obtained from
Eqs.\eqref{eq:tZgdlc}, \eqref{eq:tZtotal} by setting GDL thickness to zero:
\begin{equation}
   \tZ_{chan} = - \dfrac{4\lam \expo N_c }{\bigl(2\lam\ri\tom + (2\lam - 1)\expo\bigr) D_c}
   \label{eq:tZchan}
\end{equation}
where $N_c$ and $D_c$ are given in Appendix.
The characteristic frequency $f_{chan}$ of channel impedance
depends almost linearly of the cell current density and
on the air flow stoichiometry (Figure~\ref{fig:fchan}). It is interesting
to note that the slope of $f_{chan}(J)$ increases with the
growth of stoichiometry $\lam$ (Figure~\ref{fig:fchan}a), although
the magnitude of $\tZ_{chan}$ rapidly decreases with $\lam$.

Setting in Eq.\eqref{eq:tZchan} $\tom=0$, we find the static resistivity due to channel;
in the dimension form this resistivity reads
\begin{equation}
    R_{chan} = \dfrac{b}{J}\left(\dfrac{2}{(2\lam - 1)\ln(1 - 1/\lam)}
    - \lam\lnl{1 - \dfrac{1}{\lam}} \right)
    \label{eq:Rchan}
\end{equation}
Eq.\eqref{eq:Rchan} is independent of parameter $\xi$, Eq.\eqref{eq:xilam};
however, $\xi$ affects
the shape of impedance \eqref{eq:tZchan}. As $\xi$ contains the product $\Cdl\lcat$,
the shape of $\tZ_{chan}$ appears to be dependent of the CCL thickness.
Though for typical PEMFC parameters this dependence is weak, we should stress that
impedance of oxygen transport layers (GDL and channel) depend
on the double layer capacitance in the electrode.

\section{Conclusions}

An analytical model for impedance of oxygen transport in the GDL and channel of a PEM fuel
cell cathode is developed. The model is based on 1d+1d--coupled
oxygen mass transport equations along the channel and through the GDL. The oxygen and proton
transport through the cathode catalyst layer are assumed to be fast. Analytical
solution for the GDL+channel impedance $Z_{gdlc}$ is obtained.
The Nyquist spectrum of $Z_{gdlc}$ consists of two arcs,
of which the left arcs corresponds to oxygen transport in the GDL and
the right (LF) arc is due to oxygen transport in the channel.
The characteristic frequency of the channel arc depends linearly on the oxygen stoichiometry.
The shape of the  GDL arc differs from the Warburg finite--length
impedance due to effect of double layer charging in the catalyst layer attached to the GDL.
For typical PEMFC parameters, the characteristic frequency of the GDL arc $f_{gdl}$ is close to
the Warburg finite--length frequency. However, for larger CCL thickness the non--Warburg
factor strongly shifts $f_{gdl}$ to lower values.
At low catalyst layer thickness, the high--frequency feature in the spectrum of imaginary part of
the GDL impedance forms. This effect is particularly important for identification
of DRT peaks in low--Pt cell spectra, as low--Pt cells typically differ from the high--Pt cells
by the catalyst layer thickness only.

\appendix

\section{Equations for the factors $N_c$ and $D_c$ in Eq.\eqref{eq:tZchan}}

\begin{multline}
    N_c  = \lam^2\tJ \left(\expot + \left(\tJ + \ri\tom(1 +\xi^2)\right) \expo - \xi^2\tom^{2}\right) \\
    \times\lexp{\dfrac{ - \expo - \ri \tom\xi^2}{\lam  \tJ}} + \lam\left(\expo - \lam  \tJ + \ri\tom\xi^2\right) \\
    \times \left(\expot + \left(\tJ + \ri\tom(\xi^2 + 1)\right)\expo - \xi^2 \tom^2\right) \\
    - \left( \expo + \ri\tom\xi^2\right)^{2}\expo/2
    \label{eq:Nc}
\end{multline}

\begin{multline}
    D_c = 2\lam^2\tJ\left(\expot + \left(\tJ + \ri\tom(1 + \xi^2)\right) \expo  - \xi^2\tom^2\right) \\
    \times\lexp{\dfrac{ - \expo - \ri \tom\xi^2}{\lam  \tJ}}\expo \\
    - 2 \lam\xi^6\tom^{4} + \ri\left(2 \lam\xi^2 - \xi^2 + 4 \lam \right) \xi^{4} \expo \,\tom^{3} \\
    + 2 \xi^2 \left(\left(2 \lam\xi^2 - \xi^2 + \lam \right) \expo + \lam^2\tJ\right)\expo\tom^{2} \\
    - \ri\left(\left(2 \lam - 1\right)\xi^2\expo + 2\lam\tJ\left(\lam\xi^2 - \xi^2 + \lam\right)\right)\expot \tom \\
    - 2\lam\tJ\left(\left(\lam - 1\right)\expo + \lam\tJ\right)\expot
    \label{eq:Dc}
\end{multline}



\vspace*{1em}

\centerline{\Large\bf Nomenclature\\[1em]}

\small

\begin{tabular}{ll}
	$\tilde{}$   &  Marks dimensionless variables                             \\
	$b$          &  ORR Tafel slope, V                                        \\
    $\Cdl$       &  Double layer volumetric capacitance, F~cm$^{-3}$            \\
	$c_1$        &  Oxygen molar concentration  \\
                 & at the CCL/GDL interface, mol~cm$^{-3}$      \\
    $c_b$        &  Oxygen molar concentration in the GDL, mol~cm$^{-3}$      \\
    $c_h$        &  Oxygen molar concentration in the channel, mol~cm$^{-3}$      \\
	$\cref$      &  Reference (inlet) oxygen concentration, mol~cm$^{-3}$      \\
    $D_b$        &  Oxygen diffusion coefficient in the GDL, cm$^2$~s$^{-1}$    \\
	$F$          &  Faraday constant, C~mol$^{-1}$                            \\
    $f$          &  Characteristic frequency, Hz                              \\
  	$i_*$        &  ORR volumetric exchange current density, A~cm$^{-3}$          \\
    $\ri$        &  Imaginary unit                                            \\
	$j$          &  Local proton current density along the CCL,  A~cm$^{-2}$     \\
    $\jlim$      &  Limiting current density  \\
                 &  due to oxygen transport in the GDL, Eq.\eqref{eq:jlim} A~cm$^{-2}$     \\
	$j_0$        &  Local cell current density, A~cm$^{-2}$                   \\
    $l_b$        &  GDL thickness,  cm                                       \\
	$\lcat$      &  CCL thickness,  cm                                       \\
    $t$          &  Time, s                                                   \\
    $t_*$        &  Characteristic time, s, Eq.\eqref{eq:tast}                \\
	$x$          &  Coordinate through the cell, cm                           \\
    $Z$          &  Local impedance, Ohm~cm$^2$                               \\
    $Z_{gdlc}$   &  GDL+channel impedance, Ohm~cm$^2$                      \\
    $Z_{tot}$    &  Total cathode side impedance, including \\
                 &  faradaic one, Ohm~cm$^2$ \\
    $z$          &  Coordinate along the cathode channel, cm  \\[1em]
\end{tabular}

{\bf Subscripts:\\}

\begin{tabular}{ll}
	$0$      & Membrane/CCL interface \\
	$1$      & CCL/GDL  interface     \\
    $b$      & In the GDL \\
    $gdl$    & GDL   \\
    $gdlc$   & GDL+channel \\
    $f$      & faradaic     \\
    $h$      & Air channel             \\
    $W$      & Warburg \\[1em]
\end{tabular}

{\bf Superscripts:\\}

\begin{tabular}{ll}
	$0$      & Steady--state value \\
	$1$      & Small--amplitude perturbation \\[1em]
\end{tabular}

{\bf Greek:\\}

\begin{tabular}{ll}
    $\eta$              &  ORR overpotential, positive by convention, V     \\
    $\lam$              &  Air flow stoichiometry, Eq.\eqref{eq:xilam}      \\
    $\mu$               &  Dimensionless parameter, Eq.\eqref{eq:mu}        \\
    $\xi$               &  Dimensionless parameter, Eq.\eqref{eq:xilam}      \\
    $\phi$              &  Dimensionless parameter. Eq.\eqref{eq:phipsi} \\
    $\psi$              &  Dimensionless parameter. Eq.\eqref{eq:phipsi} \\
    $\omega$            &  Angular frequency of the AC signal, s$^{-1}$
\end{tabular}

\end{document}